\documentstyle[multicol,prl,aps,epsf,psfig,epsfig]{revtex}
\begin{document}

\title
{Earthquake aftershock networks generated on Euclidean spaces of different
fractal geometry}
\author
{Kamalika Basu Hajra and Parongama Sen}
\address
{
Department of Physics, University of Calcutta,
    92 Acharya Praffulla Chandra Road, Kolkata 700009, India. \\
}
\maketitle
\begin{abstract}

According to some recent analysis (M. Paczuski and M. Baiesi, Phys. Rev. E 
{\bf 69}, 066106, 2004 \cite{maya1}) of earthquake
data, aftershock epicenters can be considered to represent the nodes of a network 
where the linking scheme depends on several factors. In the present paper
 a model 
network of earthquake aftershock epicenters is proposed based on this 
scheme and studied on fractals of different dimensions.
The various statistical features of this network, like degree, 
link length, frequency and correlation distributions 
are evaluated and compared to the observed data. The results are also found to
be independent of the fractal geometry.
\end{abstract}



\begin{multicols}{2}

   An earthquake is a complicated spatio-temporal phenomenon \cite{eq}
 which exhibits complex correlation in space, time as well as magnitude.
 Physically, earthquakes occur when the convective motion in
 the mantle cause sudden rupture
and deformation of certain parts of the earth's crust that result in the
 emanation of energy in the form of seismic waves. In other words, earthquakes
 result from the interaction between stress concentration and fluid flow and 
have recently been the subject of considerable interest in various fields 
including statistical physics because of its unique statistical features.\\

 Depending on the relative magnitude and position in the space-time sequence 
we can primarily classify earthquakes into three categories of events
 \cite{maya1,eq,bak1} :\\
(i) Intermediate or small amplitude precursor events that precede a main event, 
known as {\it fore-shocks}.\\
(ii) Events of relatively large magnitude and impact, known as {\it main shocks}.\\
(iii) Nearby smaller correlated events that follow a large seismic event or
 main shock, known as {\it aftershocks}.\\

Earthquakes are best understood by studying large space-time correlations of 
many events instead of  observing isolated individual events and it has been 
suggested that earthquake is a critical phenomenon \cite{bak2}. The striking
 statistical features followed by earthquakes in general are the following:\\
(a) The distribution of earthquake magnitude ($m$), is given by the 
Gutenberg-Richter (GR) law \cite{GR}:
\begin{equation}
P(m) \sim 10^{-bm}, (b \sim 1)
\end{equation}
where $P(m)$ is the number of earthquakes of magnitude $m$ in a seismic
 region.\\
(b) The short time temporal correlation between earthquakes follows the 
Omori law \cite{omori}, which states that following a main event, the frequency
 of  a sequence of aftershocks decays with time $t$ as:
\begin{equation}
 N(t) \sim t^{-\alpha}, (\alpha \sim 1) 
\end{equation}
(c) The spatial distribution of earthquake epicenters form a fractal set with 
fractal dimension $d_f$ \cite{okubo}.\\
Several analyses of real earthquake data reveal that earthquakes are a result
 of a dynamical many-body system that reaches a stationary critical state 
characterised by spatial and temporal correlations that follow power laws 
(eq.s 1,2) without any intrinsic time or length scales and hence they are 
related to self organised critical phenomena \cite{sahimi,manna}.\\

{\it The Earthquake Network}:\\  

 In \cite{maya1} the earthquake aftershock phenomenon has been visualised
as a network.
Networks are complex web like structures comprising 
of nodes, connected by links and such structures describe a wide variety of 
natural systems and have been studied recently with growing interest \cite{BA}.
 In \cite{maya1} the nodes are
 the earthquake/aftershock epicenters and they are linked with a weightage given by the correlation
 between them. For the generation of an earthquake
 network, a metric is required which would quantify the correlation between 
two events or whether one event can be considered as an aftershock of another. 
This metric should incorporate the self similar statistical properties that 
unify earthquakes in general and is given by \cite{maya1}:\\

\begin{equation}
 n_{ij} \equiv Ctl^{d_f}{\Delta}m10^{-bm_i}
\end{equation}
where $i$ and $j$ represent events, i.e, earthquakes and/or aftershocks 
 time ordered with $i$ preceding $j$.\\ 
The various quantities appearing in eq.(3) are as follows:\\
$m_i$ = Magnitude of the $i^{th}$ event;\\
$l=l_{ij}$ = Spatial distance between two earthquake epicenters 
corresponding to the $i^{th}$ and $j^{th}$ events;\\
 $t=t_{ij}=T_j-T_i$ = Time interval between the two events, with
$T_j > T_i$;\\
$C$ = A constant depending on the overall seismicity in the region under 
consideration and\\
$n_{ij}$  = the expected number of events with magnitude within
${\Delta}m$ of $m_i$, occurring within the space-time domain bounded
by events $i$ and $j$.\\

Of all the earthquakes preceding $j$, there must be some event $i=i^{*}$ which
 is most unlikely to occur and for which $n_{ij}$ is minimum. Nevertheless 
since $i^{*}$ actually took place relative to $j$, inspite of being the most
 unlikely, so $i^{*}$ is the event most correlated to $j$. Hence the degree of 
correlation between any two earthquakes $i$ and $j$ is inversely proportional 
to $n_{ij}$ and two events $i^*$ and $j$ are linked if $n_{ij}$ is minimum for 
$i = i^*$. Thus 
a network is generated  which is 
directed in time where, the nodes  are the earthquake/aftershock epicenters,
  characterised by internal parameters which are magnitude, location and
 time of occurrence, and any two 
events are linked with a weight $n_{ij}$. In this paper our aim is to find 
out whether it is possible to construct a model which reproduces results
 comparable to the observed
behaviour of the statistical features of some real earthquake data. We have 
also studied the dependence of the results on Euclidean spaces of different 
fractal dimensions. \\ 

{\it Observed results for real earthquakes:}\\
In \cite{maya1}, the catalogue maintained by the Southern California 
Earthquake Data Center (SCEDC) has been analysed. The values of the various 
constants in eq.(3) that are used are 
$b \simeq 0.95$, $C = 10^{-9}$. \\
The value of the fractal dimension used is $d_f \simeq 1.6$ following the
result obtained in \cite{corral}.\\
The analysis led to the following observations:\\

(a) The distribution of the time intervals of the occurrence  
 of aftershocks, $\nu(t)$, shows a power law decay ($\nu(t) \sim \frac{K}{c+t}$
for $t < t_{cutoff}$; $c$ and $K$ are constant in time and $t_{cutoff}$
 depends on the magnitude $m$ of the earthquake) which is consistent
 with the Omori law \cite{omori}.\\
(b) The resulting network of earthquakes is scale-free, i.e., the out-degree 
distribution $P(k)$ follows a power law decay, and for the SCEDC data
  it is found that 
$P(k) \sim k^{-\gamma}$ with $\gamma = 2.0$.\\
(c) The link weight distribution ${\cal{N}}(n_{ij})$ exhibits a power law decay with slope 
$\simeq 1$.\\
(c) The link length is defined in \cite{maya1} as the distance between the 
epicenter of an aftershock and its linked predecessor and its distribution
${\cal{L}}(l)$ shows an approximate power law decay.\\

{\it Modelling and Results}:\\
 In our present work we simulate a network of 
earthquake aftershocks on Euclidean spaces with different
 values of the fractal dimension $d_f$. These are:\\
(i) Continuous two dimensional Euclidean space ($d_f = 2.0$)\\
(ii) Percolation cluster on two dimensional lattice ($d_f = 1.89$)\\
(iii) Backbone of a two dimensional percolation cluster ($d_f = 1.6$)\\
(iv) Elastic backbone of a two dimensional percolation cluster ($d_f = 1.10$)\\
After the generation of the network, we have studied the various statistical 
features, viz., time or frequency distribution, degree distribution,
 correlation distribution and link-length 
distribution. We compare the results for 
the four different types of Euclidean spaces with the observations of real 
earthquake data and also examine whether the results
at all depend upon the fractal dimensionality of the system.\\

We have simulated the earthquake network on all four kinds of 
Euclidean spaces with different system sizes pertaining to 
technical limitations. An averaging over a  maximum of  $1000$
 configurations have been used in all cases. The network 
generation procedure is as follows:\\
(i) Selection of nodes: For the continuous two dimensional
 Euclidean lattice, we take a unit square space
where nodes occur randomly 
and each node is assigned coordinates $x_1,x_2$ where $0 \le x_i \le 1$.\\ 
 In order to simulate networks on percolation clusters or its backbone 
or elastic backbone, we have taken a two dimensional square lattice where the
occupation probability is equal to the percolation threshold, $p_c = 0.592746$
\cite{stauffer}. Once it is checked that a percolation cluster 
(with $d_f = 1.89$) does occur, nodes are chosen randomly on this cluster. The
same has been done for the backbone with $d_f = 1.6$ and elastic backbone
 with $d_f = 1.10$ after identifying these from the percolation cluster.\\ 
(ii) Assignment of strengths: The selected sites represent the 
earthquake/aftershock epicenters or events and 
 the epicenters  
 are assigned strengths of magnitude $m$, 
 randomly on a scale of $1$ to $10$.\\
(iii) Calculation of Distance ($l_{ij}$): When the epicenters with their 
strengths are chosen, we calculate the Euclidean distance  $l_{ij}$ between
 any two events/epicenters $i$ and $j$ where $i$ and $j$ also denote 
their time of occurrence.\\ 
(iv) Establishment of links: We start with the second event, $j = 2$ which is
obviously linked with the only available preceding event, i.e, $i = 1$. For 
$j \ge 3$, events $i = 1, 2, 3, ...., j-1$ are considered for the calculation
of the correlation $n_{ij}$ according to eq.(3) and node $i$ is connected
to  node $j$ for minimum $n_{ij}$.\\
The values of the constants used are $b = 1.0$ and $C = 10^{-9}$.
While the maximum number of points used in the
 continuous  2{\it d} Euclidean space is $n = 5000$, for the 
percolation clusters
 we have used 
lattices of  size $L \times L$ where $L$ varies from $100$ to  $800$. On the percolation clusters we have chosen from $500$ to a maximum 
of $2000$ sites as epicenters.\\

{\it Statistical properties}:\\
After the links have been established, we have evaluated the different 
distributions 
defined earlier. The results for these distributions do not show any
 significant finite size effect in any of the fractals considered.\\
 
(I) Distribution of time interval between aftershocks: First we 
observe the nature of the distribution of time interval between 
events. This should follow the Omori law\cite{omori}, according to which
 the distribution
 should have a power law decay with an exponent $\sim 1.0$. We in fact obtain a 
power law fall with the exponent close to $1.0$ for all the four fractals (Fig. 1). For the
 elastic backbone a cutoff region is observed beyond $t \sim 100$.\\ 
\begin{center}                                                                  \begin{figure}
\includegraphics[clip,width= 4cm,angle=270]{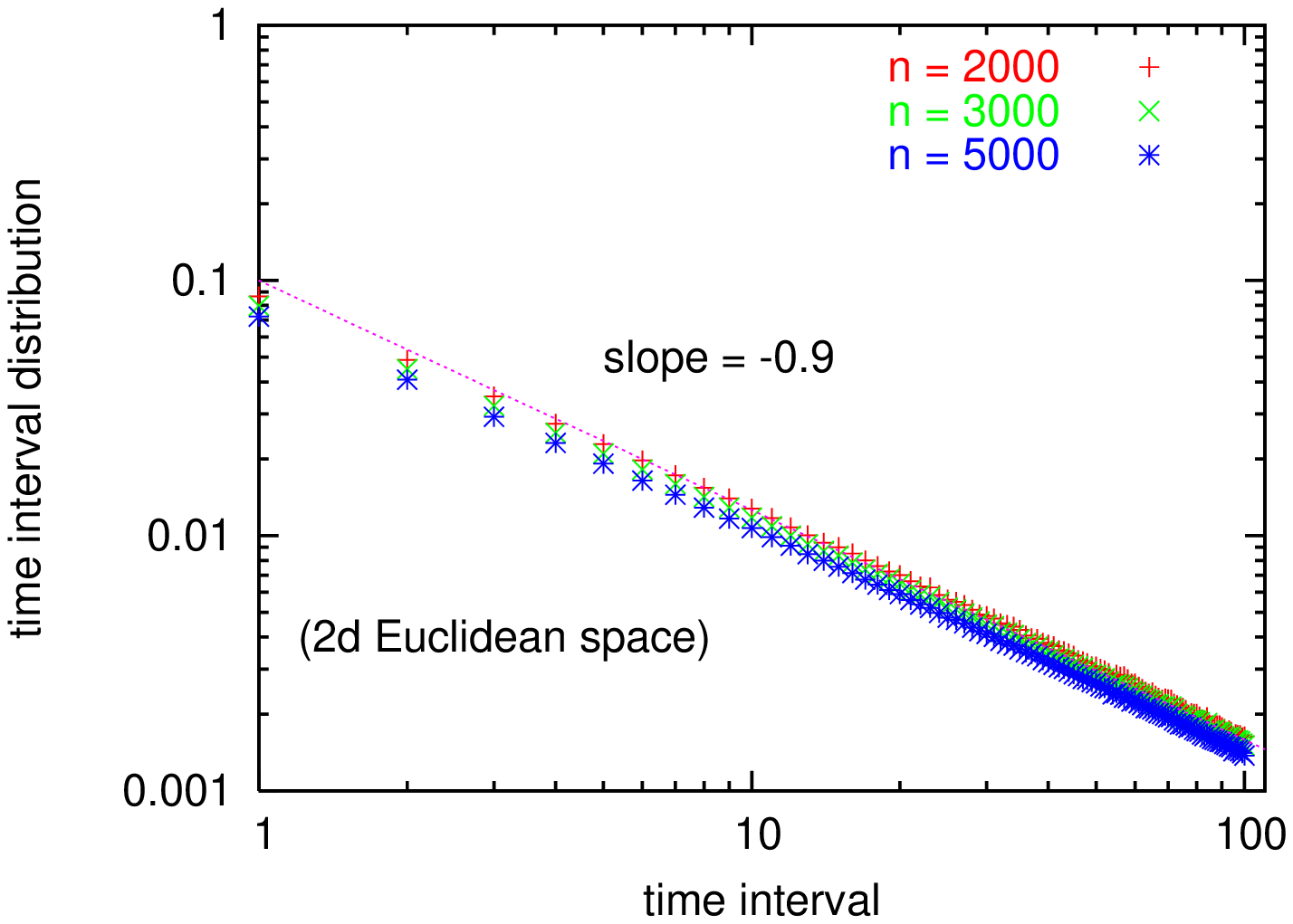}
\includegraphics[clip,width= 4cm,angle=270]{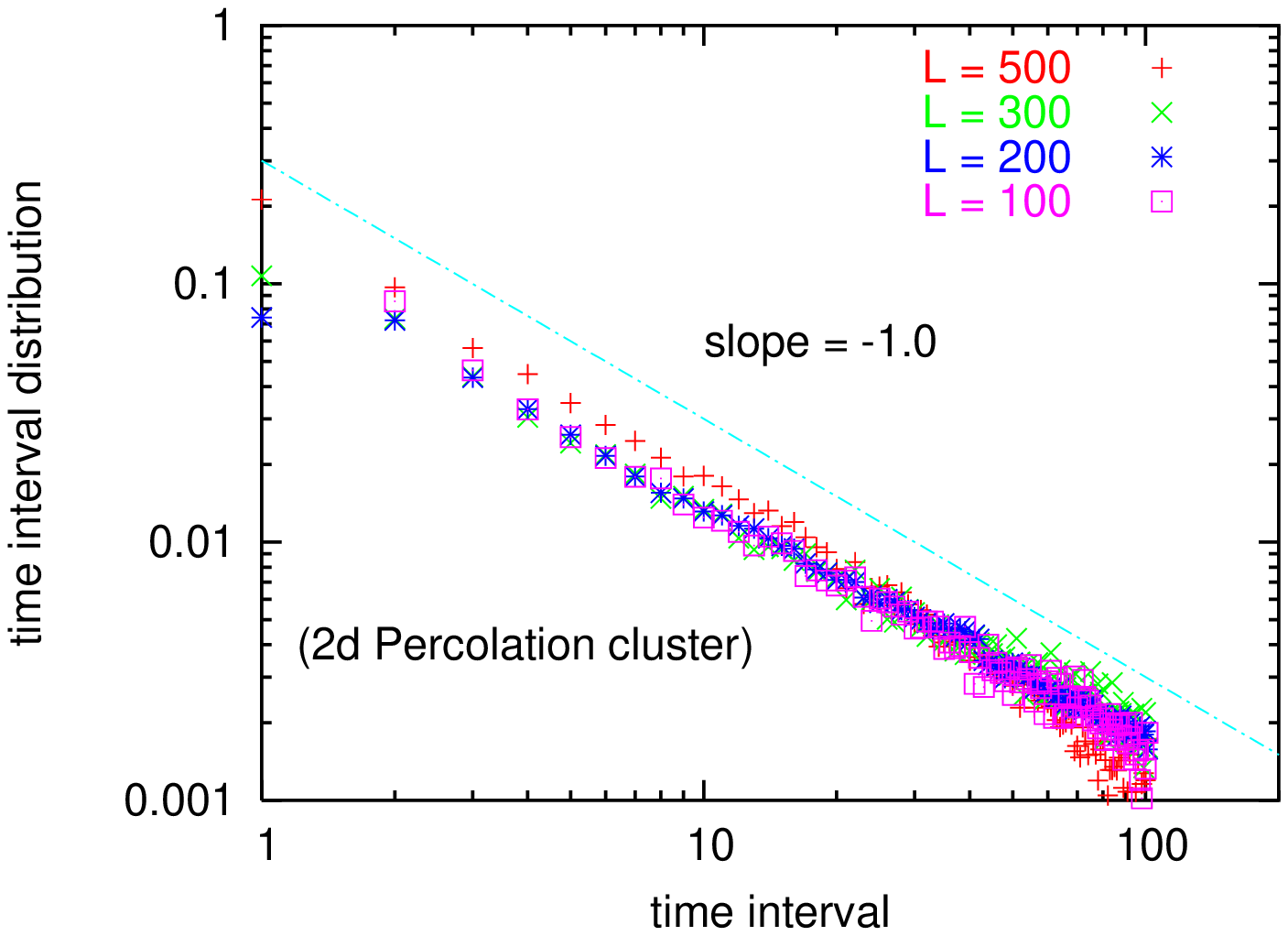}
\includegraphics[clip,width= 4cm,angle=270]{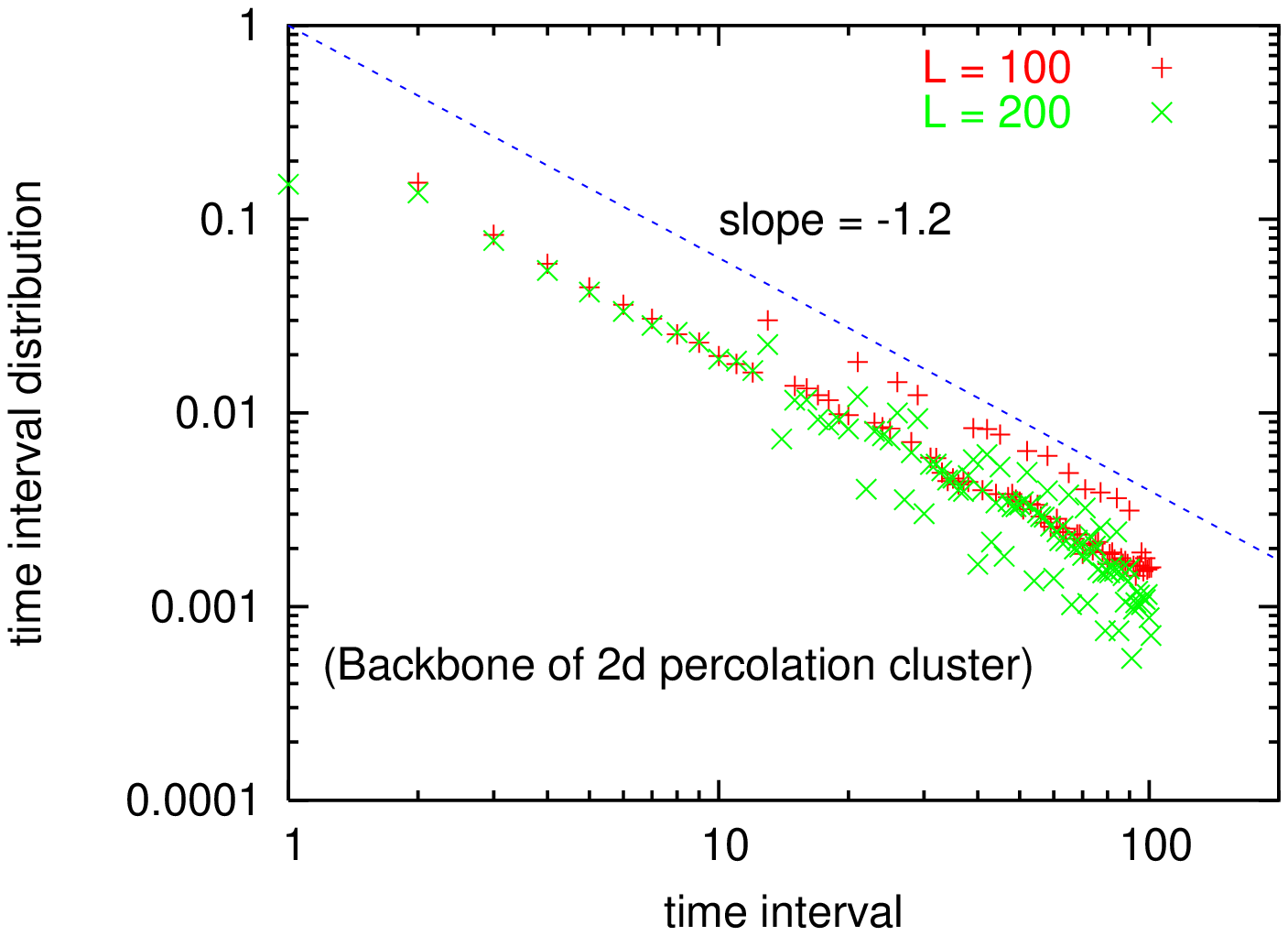}
\includegraphics[clip,width= 4cm,angle=270]{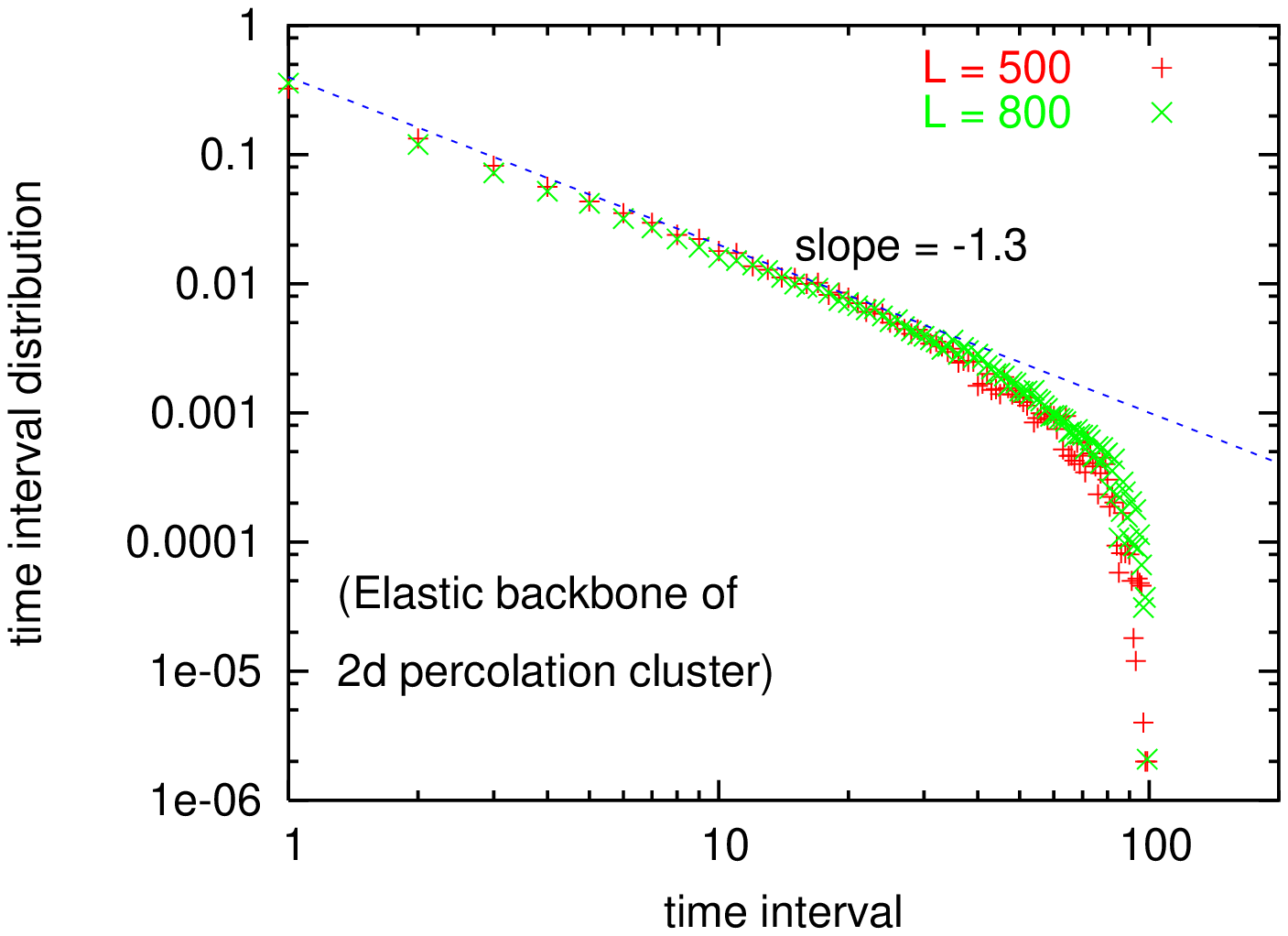}
\caption{The distribution of time interval between events. The distribution 
follows a power law decay in all cases with exponent 
close to $1.0$ which is in agreement with the Omori law. A cutoff is observed 
beyond $t \ge 100$ for the elastic backbone.}
\end{figure}
\end{center}
(II) The out-degree distribution:  The earthquake network is a directed network
\cite{maya1,BA} 
where the links are directed in time. The in-degree, i.e, the number of 
incoming links for any node is one, while the out-degree $k$ of a node gives
the number of aftershocks generated by that node. We observe that the number 
of nodes with out-degree $k$ follows a power law decrease,
 $P(k) \sim k^{-\gamma}$ 
with $\gamma \sim 2$,
  for all four types of Euclidean spaces
 in agreement with the observations of \cite{maya1}. 
Although the curves (Fig. 2) show occassional kinks, but an approximate 
straight line can be fitted for all the four cases.\\
\begin{center}
\begin{figure}
\includegraphics[clip,width= 4cm,angle=270]{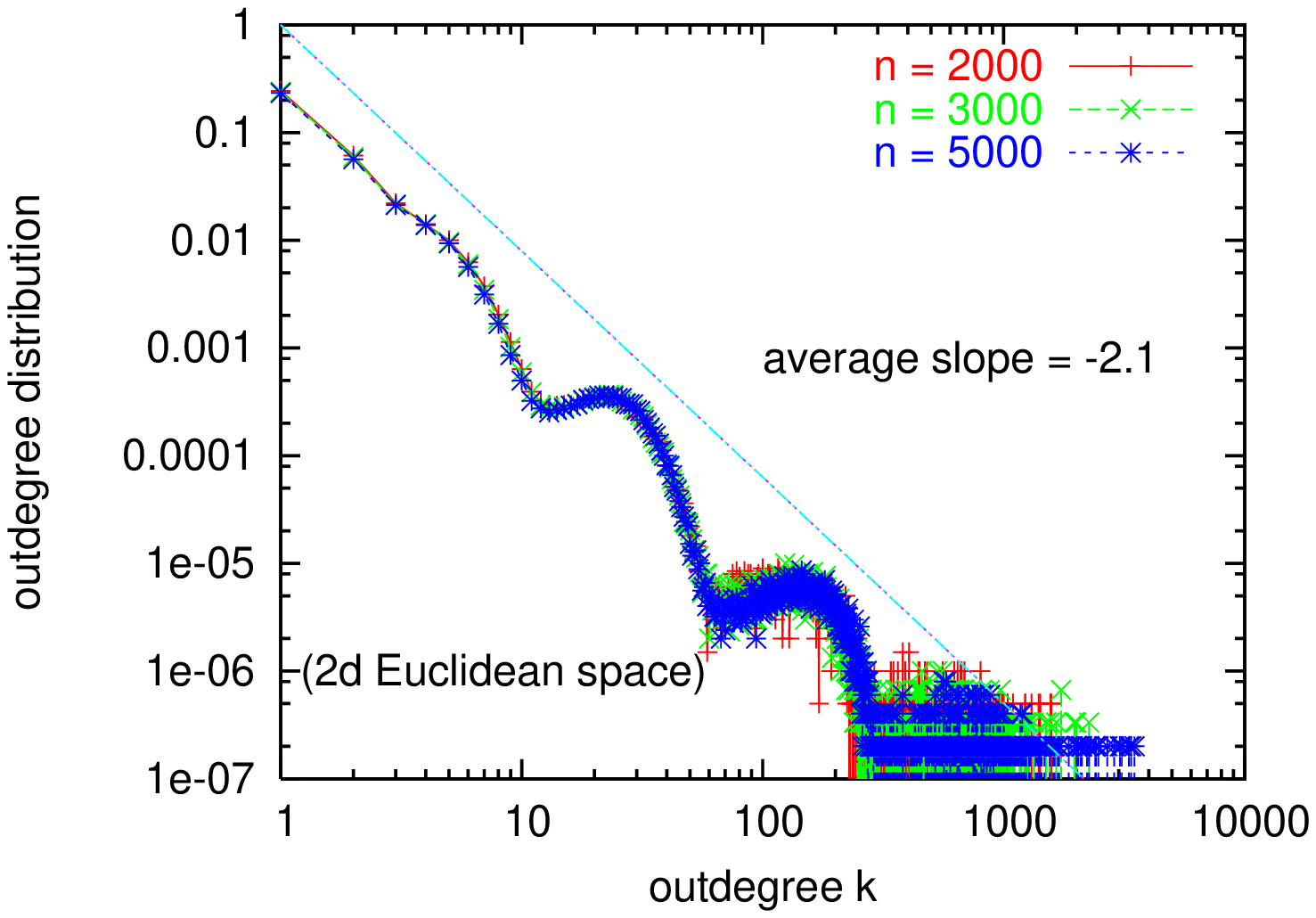} 
\includegraphics[clip,width= 4cm,angle=270]{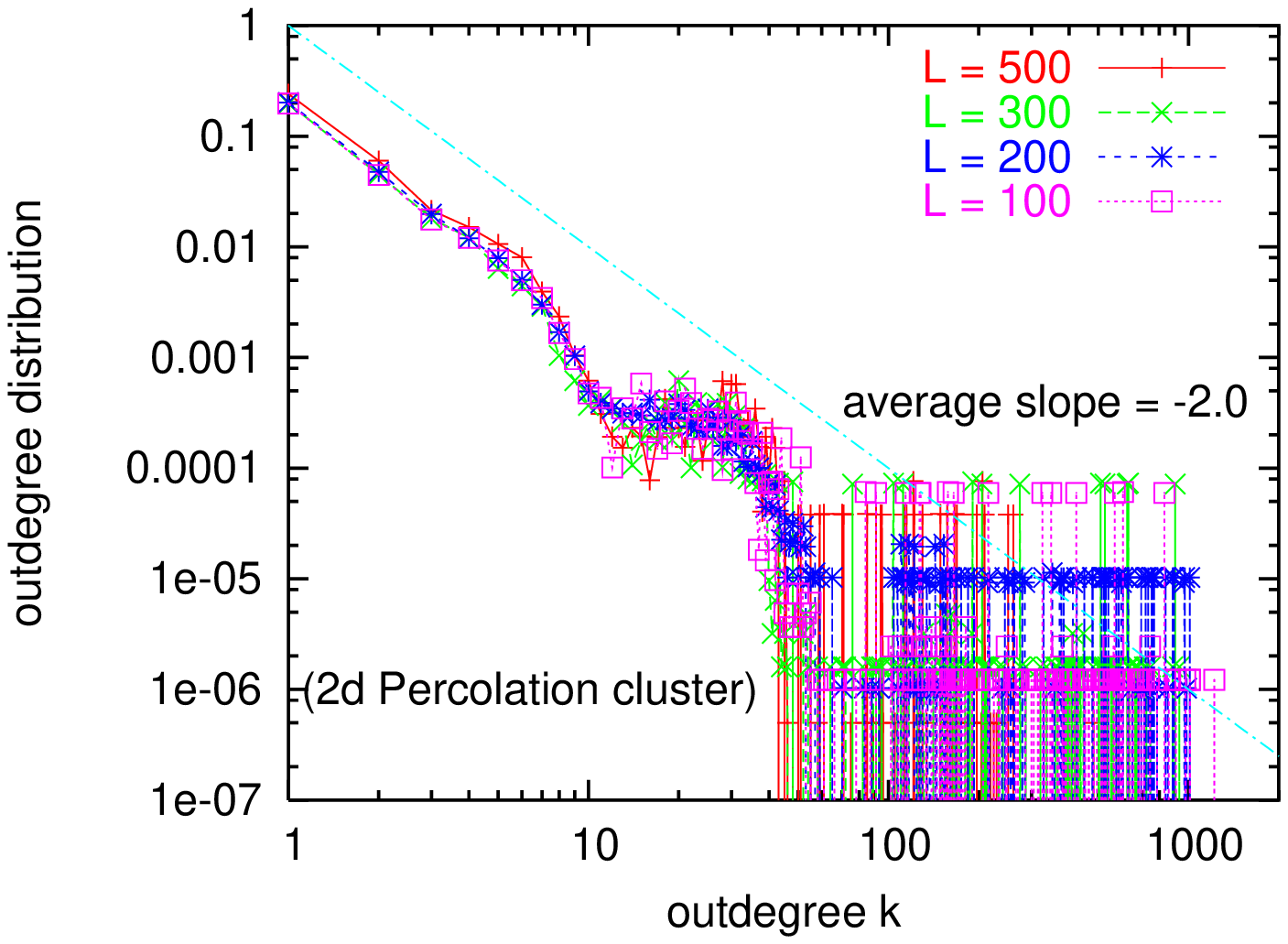} 
\includegraphics[clip,width= 4cm,angle=270]{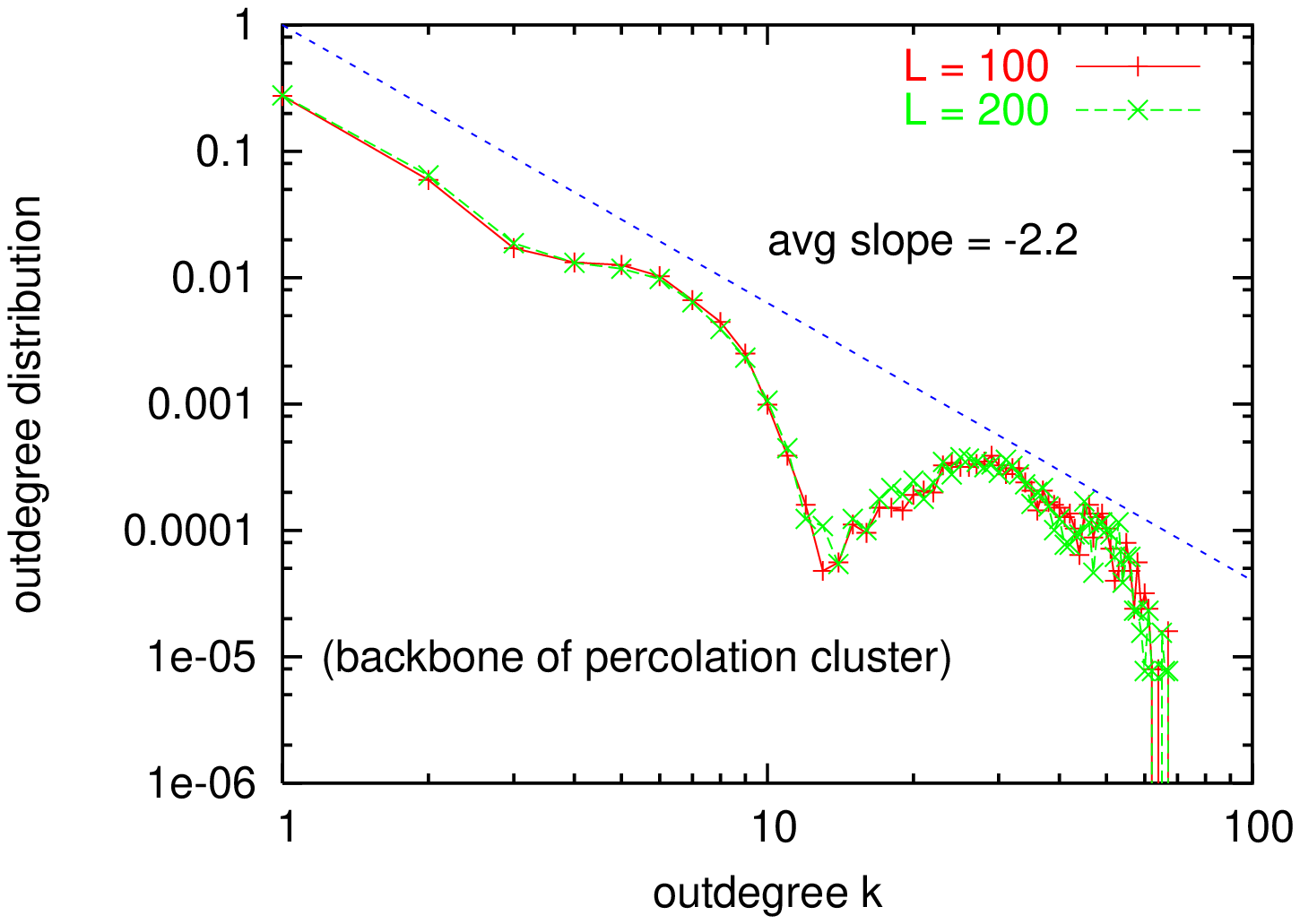} 
\includegraphics[clip,width= 4cm,angle=270]{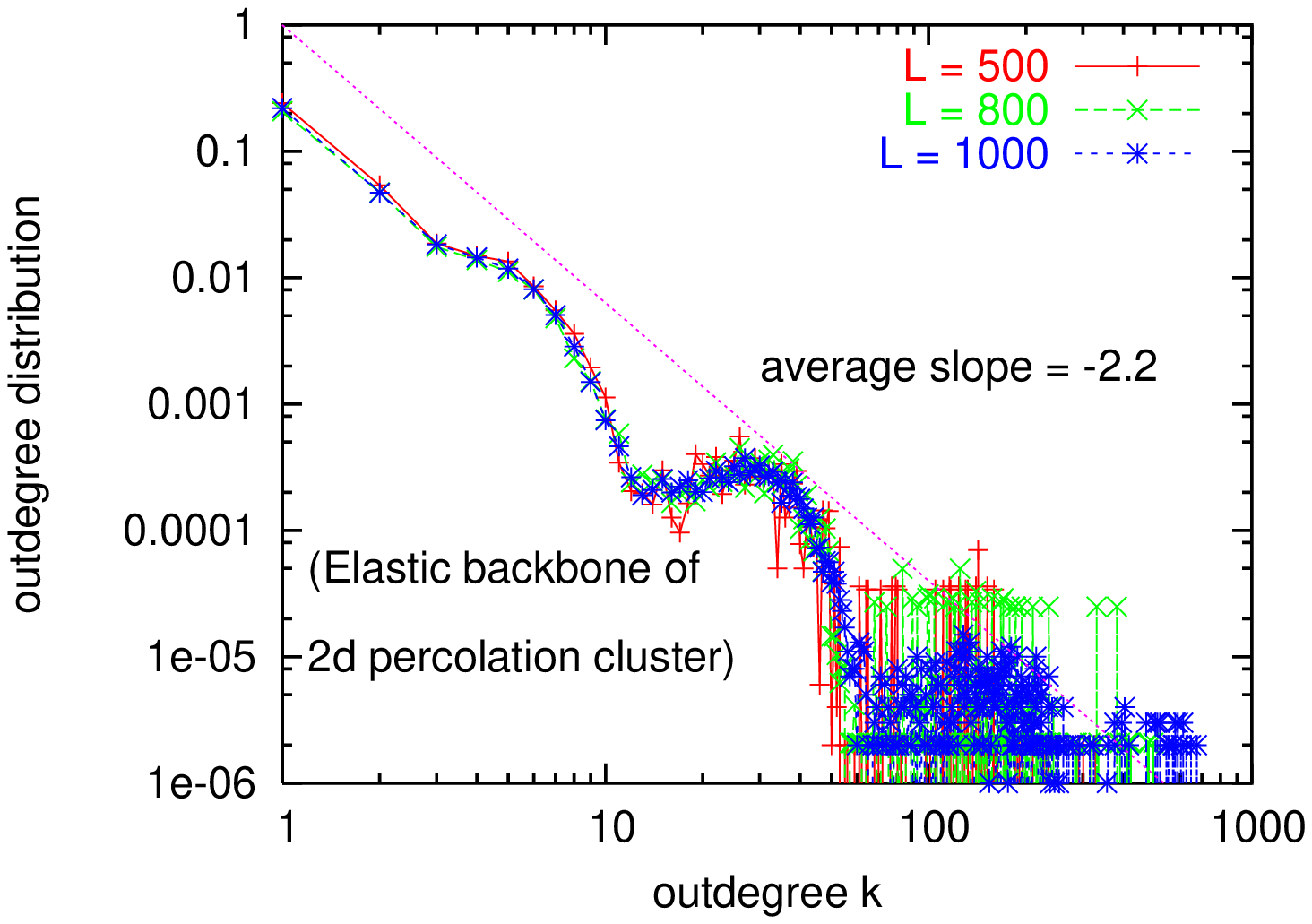} 
\caption{The out degree distribution for earthquake lattice simulated on the 
four kinds of spaces lattice as mentioned in the figures above. In each case
 it closely follows a power law decrease with average slope around $2.0$.} 
\end{figure}
\end{center}

(III) Correlation Distribution:  The plot (Fig. 3) of the correlation distribution 
${\cal{N}}(n_{ij})$ also has a power law decay, with a slope $\sim 2.6 \pm 0.2$
which deviates considerably from the observed value in \cite{maya1} (this issue
has been discussed later).
A cutoff region exists here for 
all the cases.\\
\begin{center}
\begin{figure}
\includegraphics[clip,width= 4cm,angle=270]{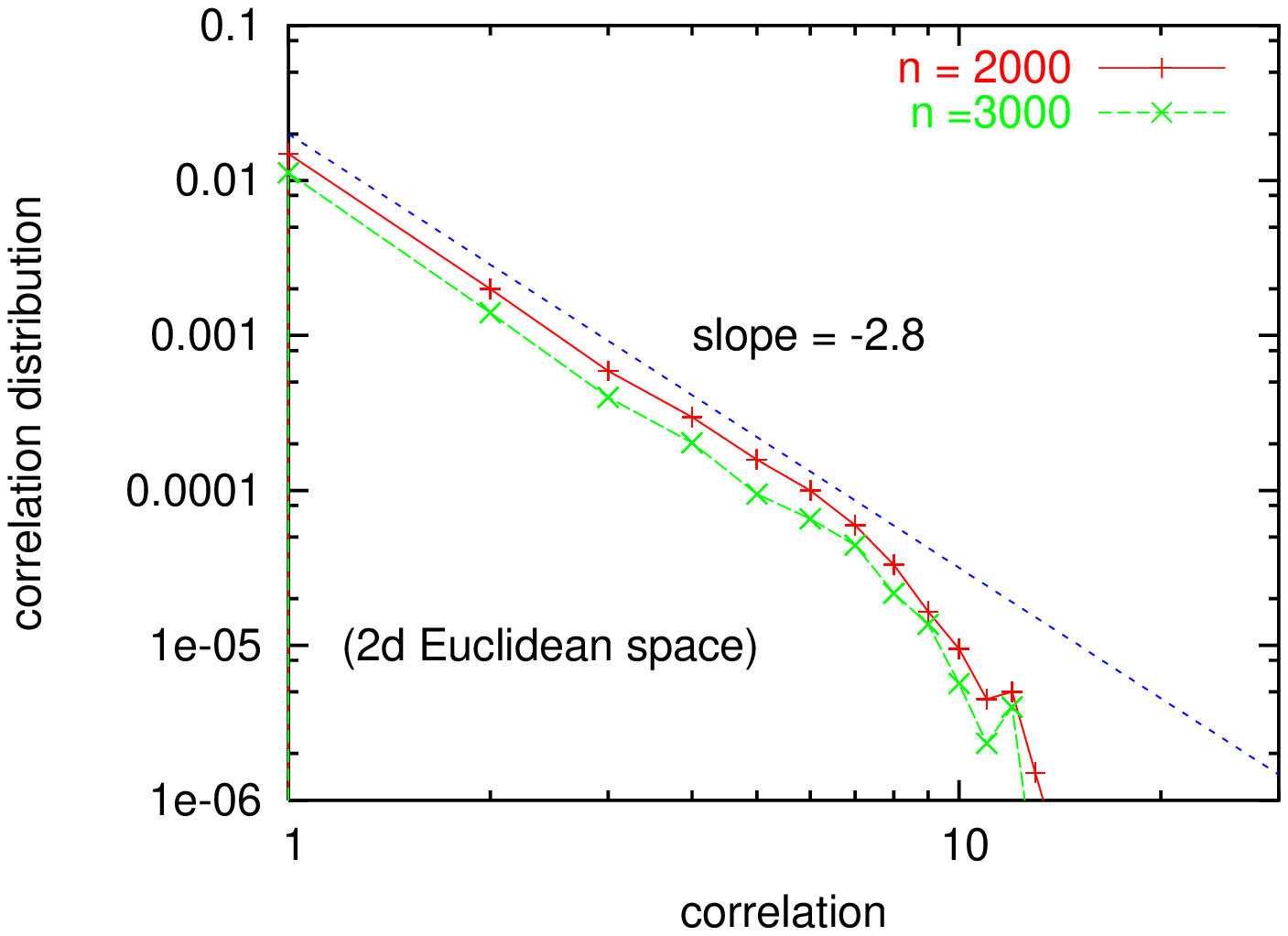}
\includegraphics[clip,width= 4cm,angle=270]{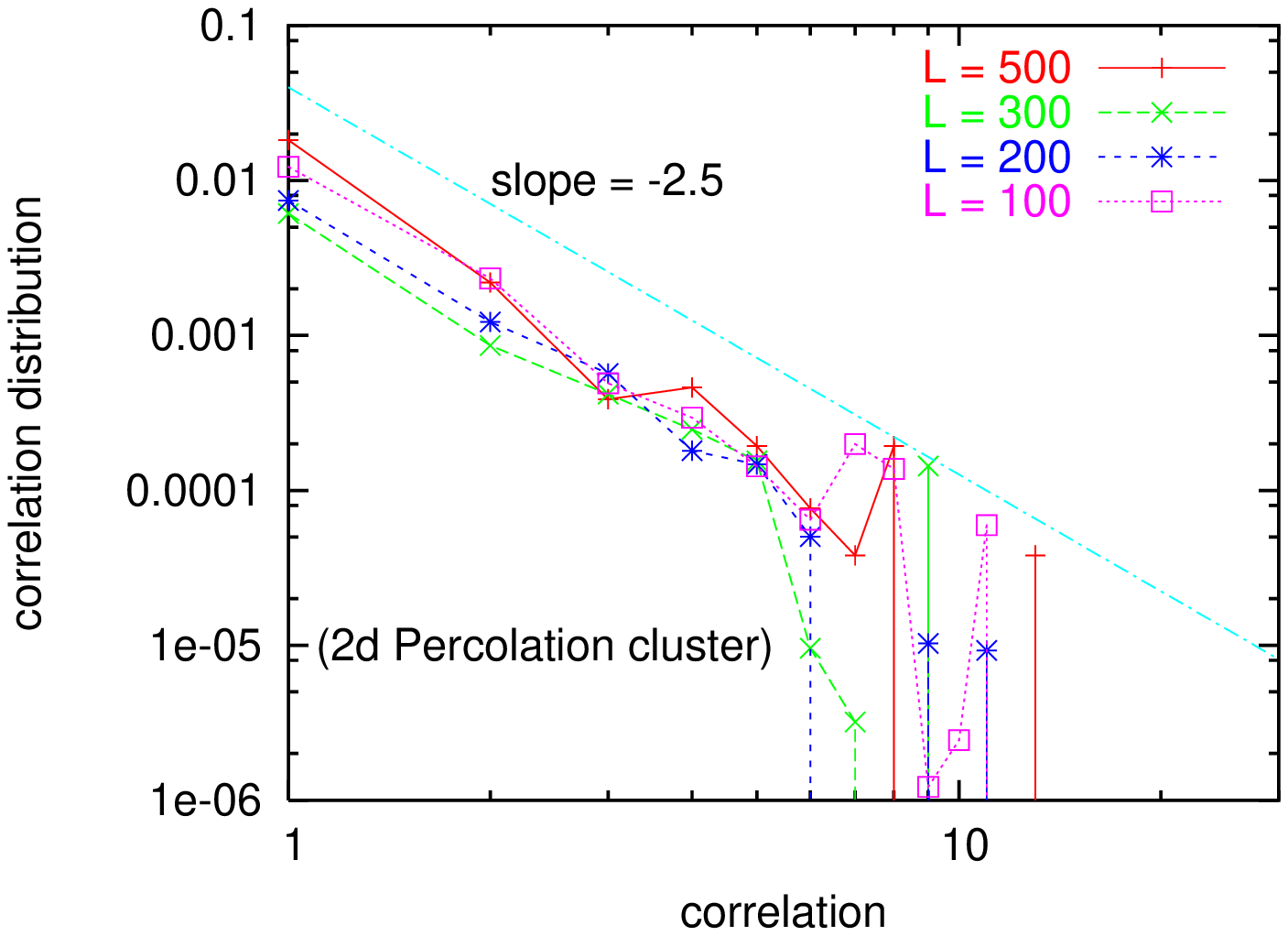}
\includegraphics[clip,width= 4cm,angle=270]{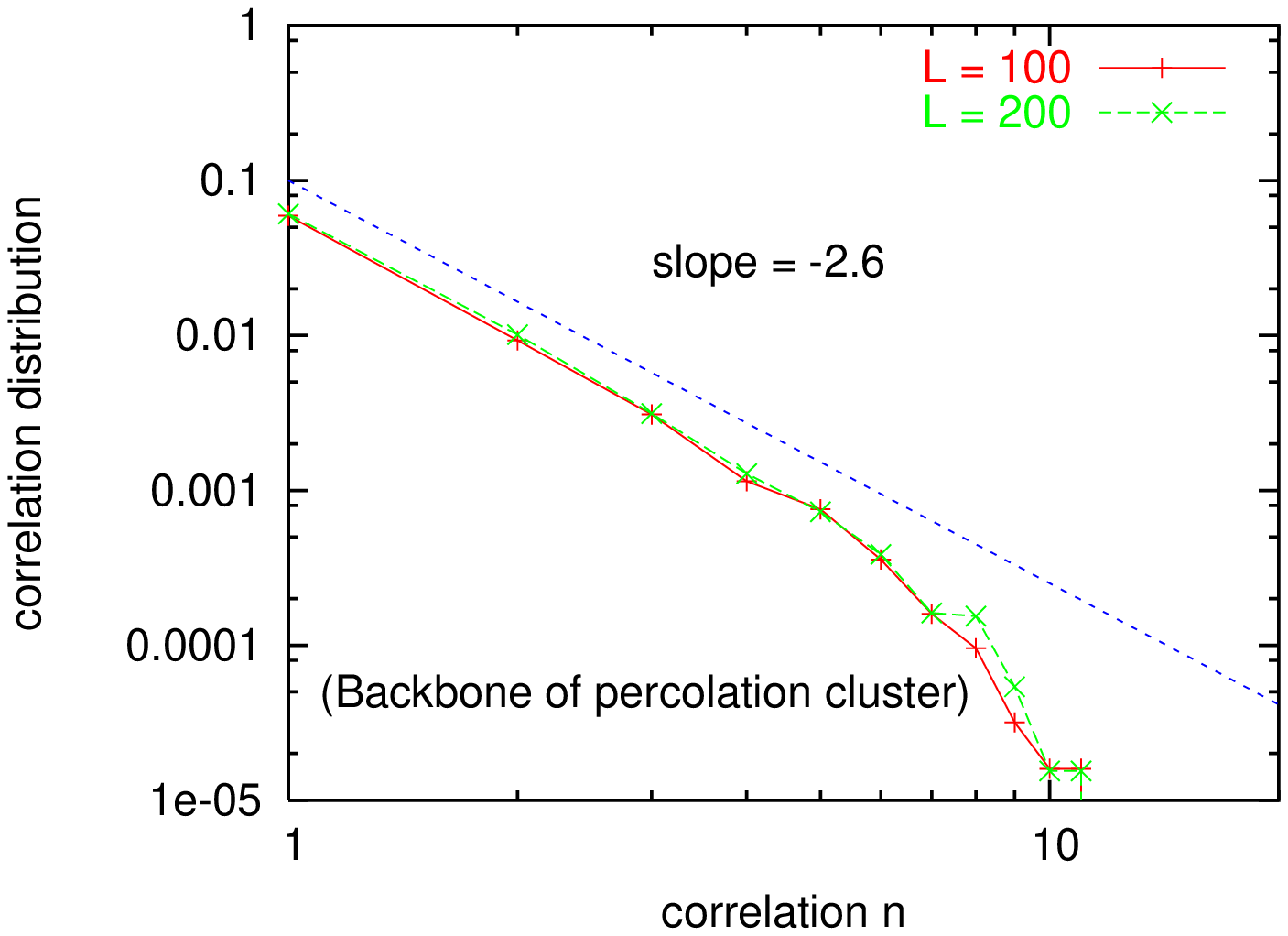}
\includegraphics[clip,width= 4cm,angle=270]{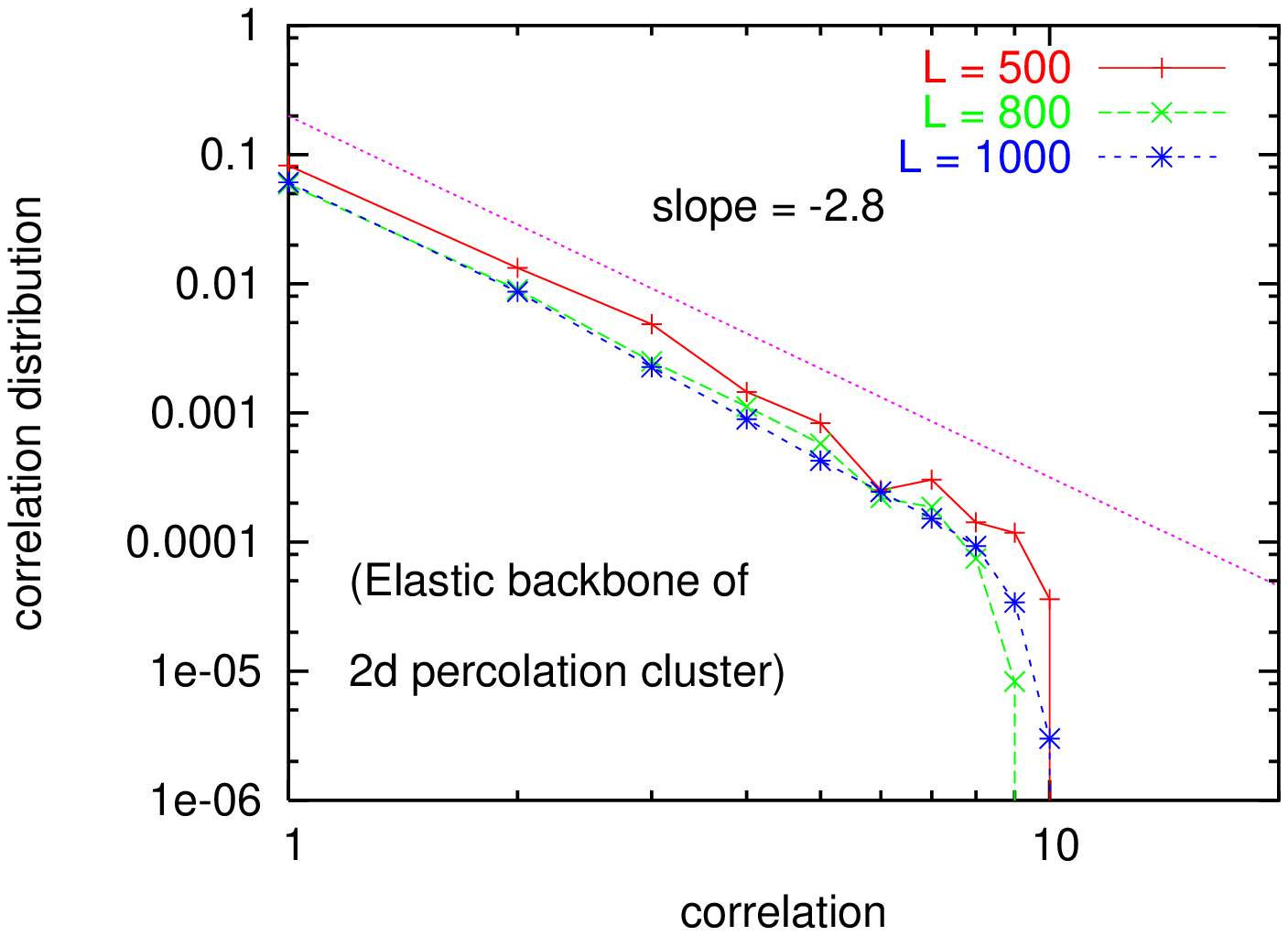}
\caption{ The plots of the correlation distribution for the four types of  
fractals. Here the plot follows a power law decay in all cases with 
slope $\sim 2.6 \pm 0.2$.}
\end{figure}
\end{center}

(IV) Link length distribution: The link-length 
distribution ${\cal{L}}(l_{ij})$ for the network 
grown on the different spaces does not 
follow a power law decay except for the elastic backbone of the two dimensional
 percolation cluster,
with slope $\sim 1.2$. For the other three cases, i.e., the continuous 
Euclidean lattice, the two dimensional percolation cluster and the backbone of
 the percolation cluster, the distributions show a power law decay with an
 exponential
cut off given by, 
 $ax^{-\rho}\exp(-{\lambda}x)$ where $a, \rho$ and $\lambda$ are constants.
The values of $\rho$($\sim 0.1$) and $\lambda$($\sim 0.1$) for the three cases
are comparable; exact values are indicated in Fig. 4.\\
 
\begin{center}
\begin{figure}
\includegraphics[clip,width= 4cm,angle=270]{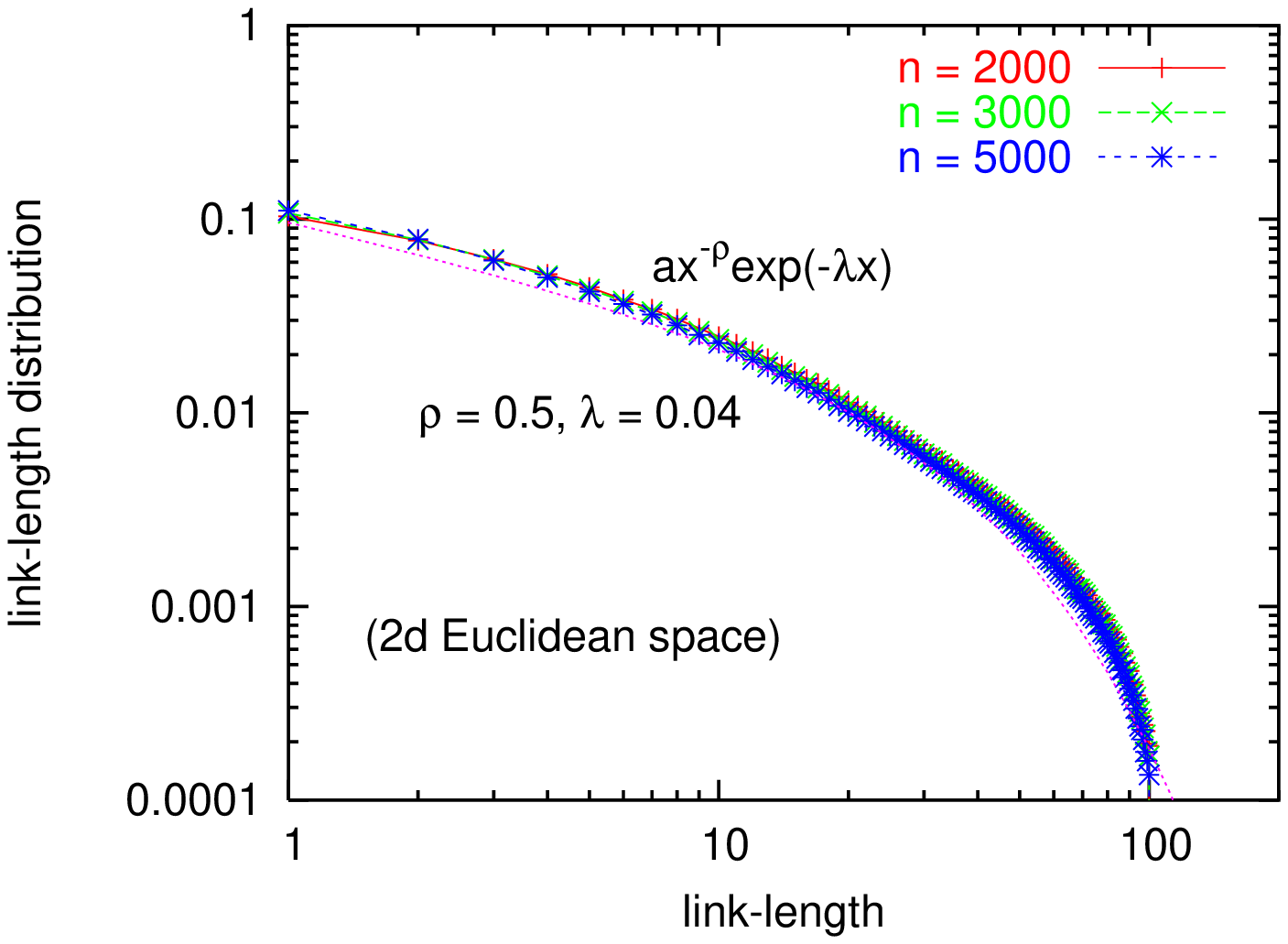}
\includegraphics[clip,width= 4cm,angle=270]{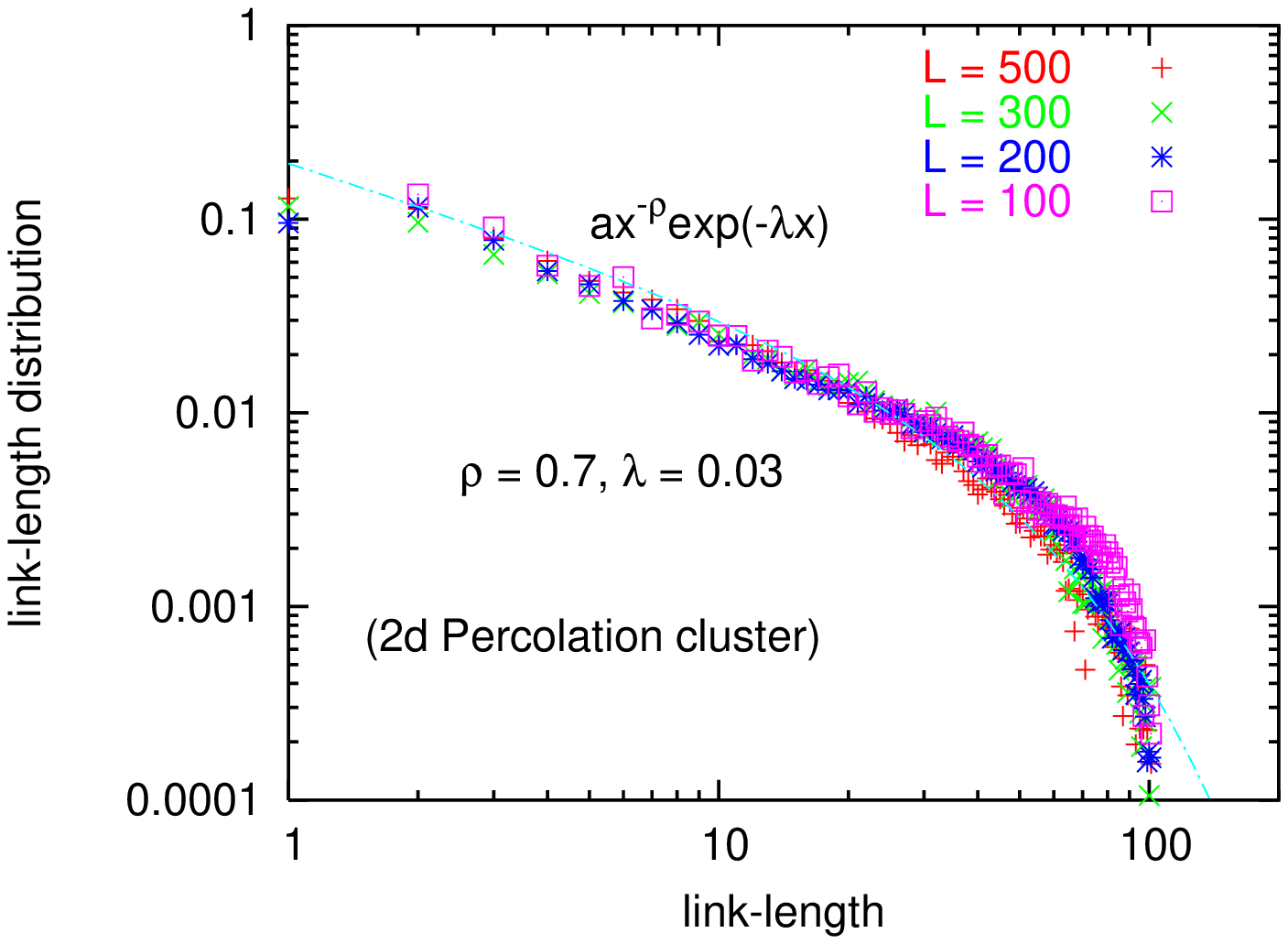}
\includegraphics[clip,width= 4cm,angle=270]{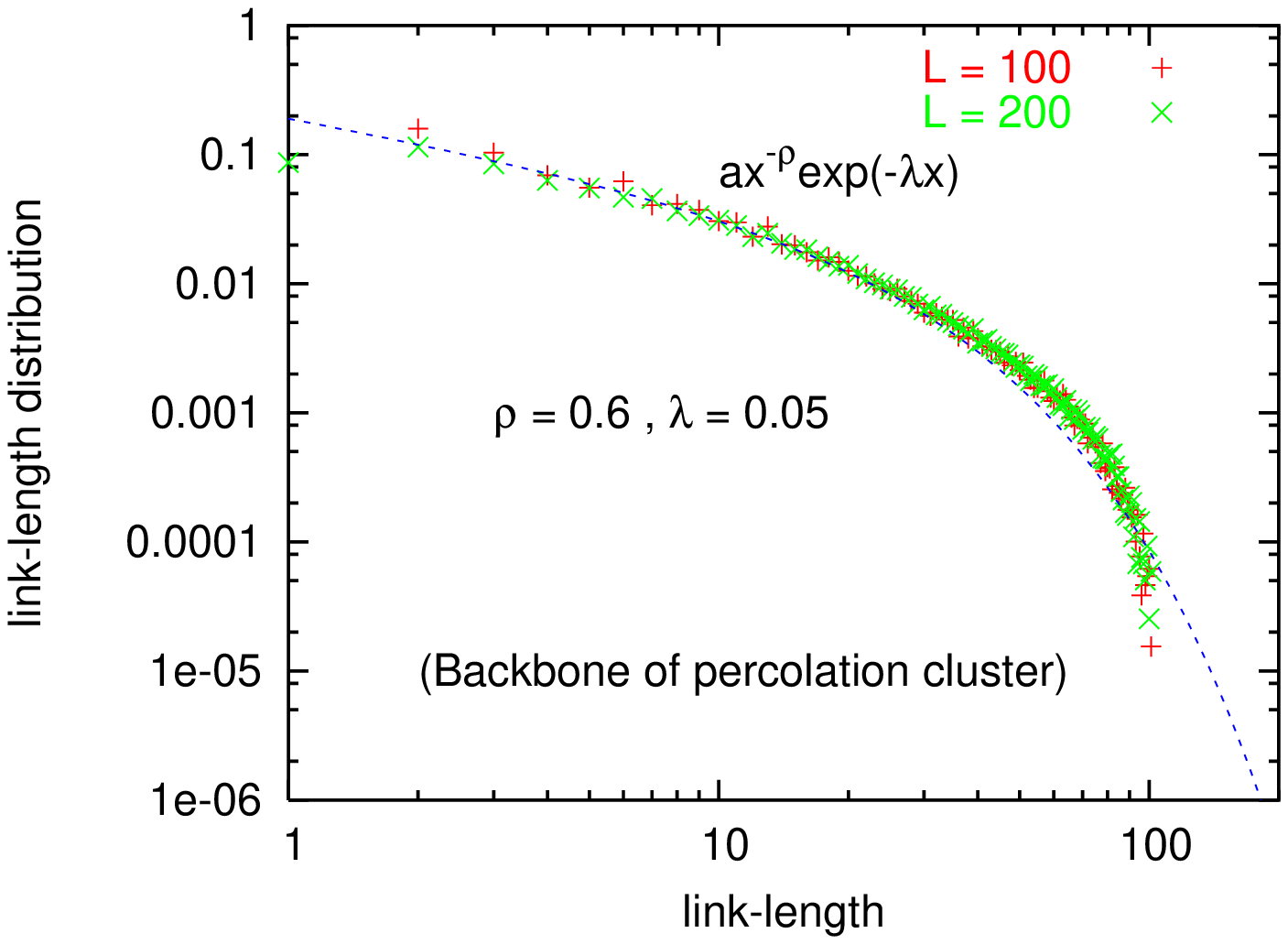}
\includegraphics[clip,width= 4cm,angle=270]{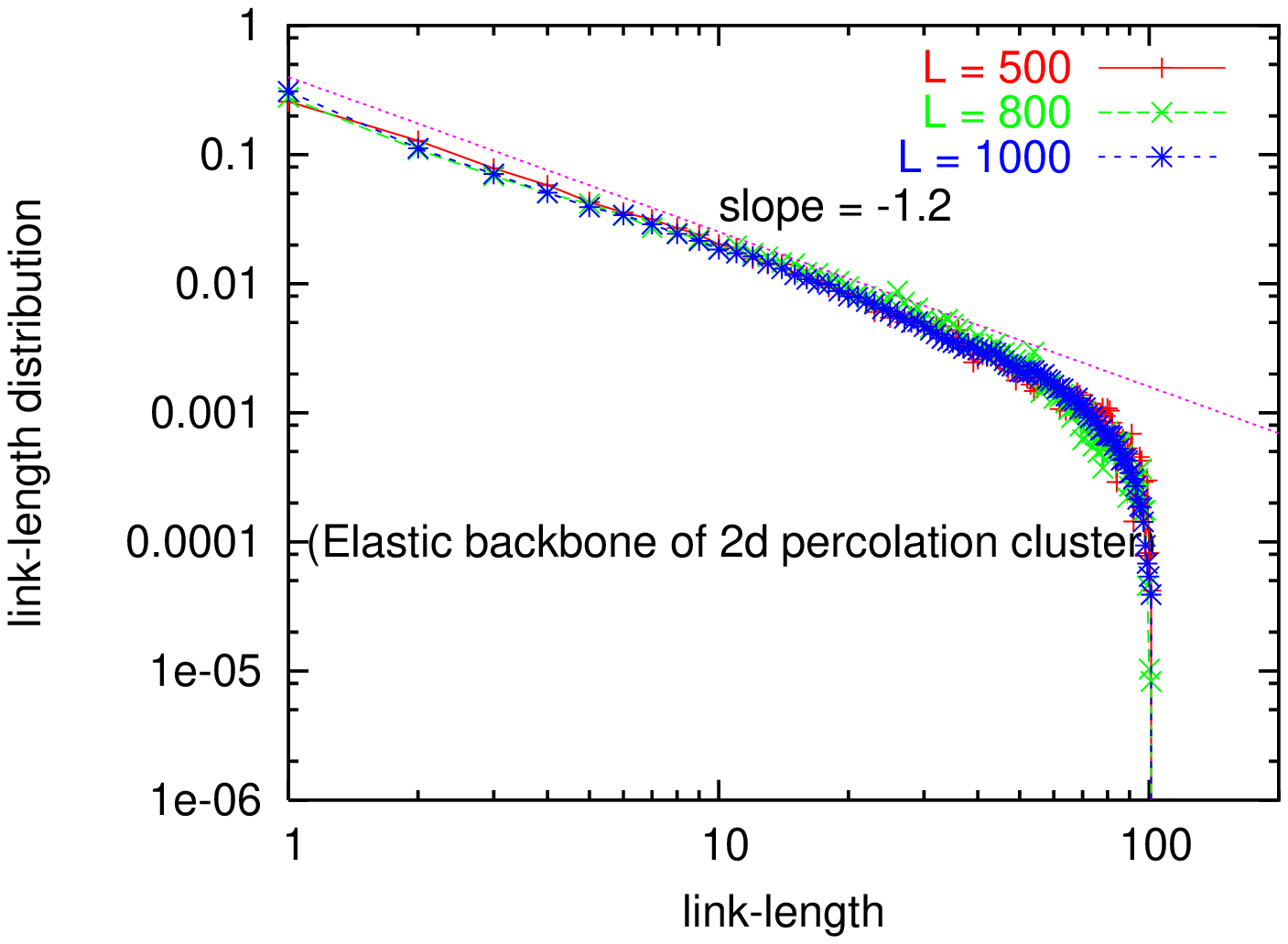}
\caption{The link length distributions for the four types of fractals.}
\end{figure}
\end{center}

To summarise the results, we have generated earthquake aftershock
 networks on Euclidean 
spaces with different fractal dimensions. The linking scheme follows the 
prescription of \cite{maya1}. We have compared the distribution functions
corresponding to the time intervals, out-degree, correlation and link-lengths 
from these networks. The results partially agree with the real data, e.g., 
the time interval distribution shows a power law decay as given by the 
Omori law, the out-degree distribution has an approximate power law decrease -
 in both cases the exponents also agree with the observed values quite well. 
The link-length distribution in \cite{maya1} is approximately a power law while
 here it seems to have an exponential cutoff.
For the correlation distributions the exponent value
deviates appreciably from the observed one. While $\nu(t)$, $P(k)$ and 
${\cal{L}}(l)$ can be estimated independently without a
 network formalism, $n_{ij}$ is a 
measure directly linked to the networking scheme. That our result for 
${\cal{N}}(n_{ij})$ does not agree quantitatively with that of the SCEDC data
\cite{maya1} may be due to the fact that epicenters have been assumed to occur 
randomly while in reality there is expected to be a correlation. However, the 
exact values of ${\cal{N}}(n_{ij})$ (which depend on the parameters chosen)
 may not be of much significance since the task is to link the epicenters with 
the minimum $n_{ij}$. The fact that we get good agreement for $\nu(t)$ and 
$P(k)$ and fairly good agreement for ${\cal{L}}(l)$ reflects this.
On the theoretical side our studies show an interesting result that
 the behaviour of these distributions are almost independent of the
 underlying spatial
 structure. We have varied the fractal dimensionality from $2$ to $1.1$ and 
we do not notice any appreciable change in the behaviour of the different 
quantities, either qualitatively or quantitatively.\\

 Acknowledgments: We thank S. S. Manna for useful comments. KBH is grateful to CSIR (India) F.NO.9/28(609)/2003-EMR-I for financial support.
PS acknowledges CSIR 03(1029)/05-EMR-II.

Email: kamalikabasu2000@yahoo.com,\\ psphy@caluniv.ac.in

\end{multicols}

\end{document}